\documentclass[prl,amsmath,amssymb]{revtex4}
\newcommand{\figurewidth}{8cm}
\usepackage{graphicx}% Include figure files
\usepackage{dcolumn}% Align table columns on decimal point
\usepackage[T1]{fontenc}
\usepackage[latin1]{inputenc}

\usepackage{bm}% bold math

\begin{document}

%\preprint{APS/123-QED}

\title{In-Situ Contacting and Current-Injection into Samples in Photoemission Electron Microscopes}

\author{L. Heyne}
\affiliation{Fachbereich Physik, Universit{\"a}t Konstanz,
Universit{\"a}tsstr. 10, D-78457 Konstanz, Germany}

\author{M. Kl{\"a}ui}\thanks{Current address: Laboratory of Nanomagnetism and Spin Dynamics, Ecole Polytechnique Federale de Lausanne (EPFL), 1015 Lausanne, Switzerland and SwissFEL, Paul Scherrer Institut, 5232 Villigen PSI, Switzerland e-mail: Mathias.Klaeui@epfl.ch}
%\email{mathias.klaeui@uni-konstanz.de}
\affiliation{Fachbereich Physik, Universit{\"a}t Konstanz,
Universit{\"a}tsstr. 10, D-78457 Konstanz, Germany}

\author{L. Le Guyader}
\affiliation{Swiss Light Source, Paul Scherrer Institut, 5232 Villigen PSI, Switzerland}

\author{F. Nolting}
\affiliation{Swiss Light Source, Paul Scherrer Institut, 5232 Villigen PSI, Switzerland}

\date{\today}

\begin{abstract}
Studying the interaction of spin-polarized currents with the magnetization configuration is of high interest due to the possible applications and the novel physics involved. High resolution magnetic imaging is one of the key techniques necessary for a better understanding of these effects. Here we present an extension to a magnetic microscope that allows for in-situ current injection into the structure investigated and furthermore for the study of current induced magnetization changes during pulsed current injection. The developed setup is highly flexible and can be used for a wide range of investigations. Examples of current-induced domain wall motion and vortex core displacements measured using this setup are presented.

\end{abstract}

%\pacs{73.23.-b, 72.25.Dc, 72.10.Fk}
%\keywords{}

\maketitle
%%%%%%%%%%%%%%%%%%%%%%%%%%%%%
%\input{defs08.tex}
%%%%%%%%%%%%%%%%%%%%%%%%%%%%%%%

\section{ I. Introduction}
The study of spin-torque phenomena resulting from the interplay of spin-polarized currents and magnetization has attracted much interest over the last years due to the exciting physics and the potential of new spin-torque based logic and storage devices \cite{Allwood2005, Parkin2008}. For instance in magnetic wires domain walls are formed at the interface of regions with opposite magnetization (domains). These domain walls can be moved by injected current pulses in the direction of the electron flow (See \cite{Klaeui2008} and references therein).

Studying current-induced magnetization switching is most often achieved by transport measurements employing the Anisotropic Magneto-Resistance effect \cite{Grollier2003,Laufenberg2006, Thomas2006,Ishida2006a} due to the measurement simplicity. However transport measurements yield only limited information, as they do not reveal the internal magnetic spin structure. This means for instance, current-induced domain wall spin structure changes can only be studied by magnetic imaging \cite{Heyne2008}.
Magnetic imaging by  X-ray Magnetic Circular Dichroism - Photo Emission Electron Microscopy (X-PEEM) is a powerful technique that is used to map the internal magnetization configurations in nanoscale structures \cite{Stoehr1993, Quitmann2008}. The emission rates of the photoelectrons created by the incident circularly polarized X-ray photons depend on the magnetic orientation inside the material with respect to the X-ray polarization \cite{Schuetz1987}. Thus imaging the spatially resolved emission rates for the two X-ray polarities yields the magnetic contrast of the sample \cite{Stoehr1993}.
However, commercial microscopes do not provide the possibility to contact the investigated structures to conduct transport measurements or to inject current.

In this paper we present a specially developed setup for the widely used ELMITEC PEEM that greatly extends the possibilities of the microscope and allows for in-situ current injection of microsecond and  nanosecond short current pulses. In addition, small magnetic fields can be applied and by gating the imaging unit, we can study the magnetic system during the pulse injection. This setup is used to image domain wall motion due to current pulses and vortex core displacements during current injection.

\section{ II. Experimental}

The PEEM employed is a commercial ELMITEC-PEEM widely used at various synchrotron sources (Type SPE LEEM III \cite{Elmitec}). % that has been adapted in the following ways to be suited for the current-injection experiments.
It offers high resolution magnetic imaging. However, one drawback in this setup is that the sample is not at ground potential but at a high voltage (HV) of about 20\,kV. This has the advantage that the magnetic lenses of the microscope are at ground potential, but complicates access to the sample, which is at 20kV. In this paper, the following modifications to the PEEM are presented that allow for a flexible in-situ current injection into the structure investigated.
\begin{itemize}
\item A special sample holder that provides four electrical contacts and allows for convenient contacting of the structure.
\item A pulse injection unit that fits into the 19" HV rack and allows for the injection of microsecond short current pulses into different structures. An additional gating unit that synchronizes the imaging with the current pulses allows for imaging during current pulse injection.
\item A fast photo diode fitted into the sample holder and triggered by a femtosecond laser pulse to generate about 3\,ns short current pulses.
\end{itemize}
Prior to these implementations, nanosecond pulse injection has not been possible at all and microsecond pulses could only be injected by switching off the high voltage, opening the HV-rack  (making sure that the HV was completely off) and injecting a current pulse using a normal pulse generator setup assuming that somehow the heating filament of the standard sample holder was connected to the sample. Our new setup explained in the following drastically reduces the workload, and the time necessary for the current injection is reduced from more than 10\,minutes to a few seconds. The next sections discuss these modifications in more detail.

\subsection{ A. Sample holder }

\begin{figure}
\includegraphics[width=\figurewidth]{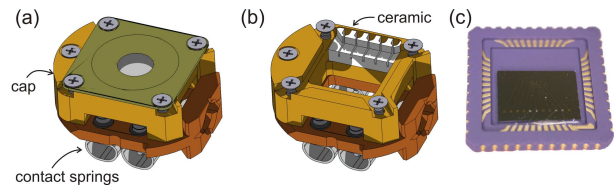}  %8cm   14
\caption[Photograph and drawing of the self-designed sample holder]{\label{figure:sampleholder}(a) Technical drawing  of the sample holder with chip carrier capability. Four flexible contacts at the bottom of the sample holder lead to the sample.  (b) Sample holder with removed cap and chip carrier showing the hollow structure to insert the magnetic coil and the ceramic part that contains the contacts for the chip carrier. (c) Image of the chip carrier with a mounted sample and wires bonded between sample and chip carrier.}
\end{figure}

The sample holder conventionally used includes a heating filament and a thermocouple, but no additional contacts.
A special sample holder design is thus required as the original holder does not provide for contacting of the sample. The new holder uses the contacts initially designed for the heating of the sample and for measuring the sample temperature and these contacts are rewired to connect to the sample and also for driving a small magnetic coil integrated into the sample holder. The sample is not directly mounted on the sample holder but on a chip carrier which allows for an easy and flexible exchange of the samples. Technical drawings of the sample holder are shown in Fig. \ref{figure:sampleholder}. In Fig. \ref{figure:sampleholder}(a) the sample holder is shown with a mounted sample and the cap.  The hollow structure of the sample holder is visible in Fig. \ref{figure:sampleholder}(b) where the cap and the chip carrier are removed. This geometry entails the possibility to insert coils into the sample holder for in-situ application of small magnetic fields. Depending on the coil geometry, either in-plane fields or out-of-plane fields of up to 200\,G can be applied. A photograph of the chip carrier with a mounted sample is shown in Fig. \ref{figure:sampleholder}(c). The sample structures are wire bonded to the chip carrier.

 To prevent discharges it is important that the sample holder contains as little insulating materials as possible to avoid charging. The ceramic part that supports the contacts connecting the chip carrier is  therefore as small as possible and shielded by metal from all sides [see Fig. \ref{figure:sampleholder}(b)]. The cap further shields the bond wires from the applied HV to further reduce the risk of discharges.

Additionally a small photo diode can be mounted on the sample holder which is required for the generation of nanosecond short current pulses (see section C).

\subsection{ B. Microsecond Pulse Injection}
Since the sample is at approximately 20\,kV with respect to the ground potential, special considerations have to be done concerning the pulse injecting unit. The original cables connecting the PEEM and the HV rack have a high frequency 3dB-cutoff at about 800kHz, which limits the minimum pulse length to a few microseconds.

\begin{figure}
\includegraphics[width=\figurewidth]{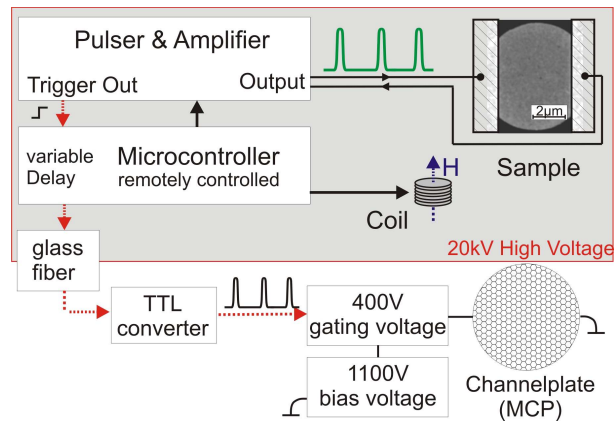}  %8cm   14
\caption[Schematic of the MCP gating]{\label{figure:gate-setup}Schematic overview of the microsecond pulse injection experiment and the MCP gating experimental setup. A pulse generator in combination with a amplifier is used to inject current pulses into the sample. A trigger signal synchronized to the current pulses is passed to the external gating setup to make sure that the imaging unit is switched on and off accordingly to the current pulse timing.

The pulse generator and the trigger signal are controlled by an external computer via a glass fiber cable and a  microcontroller. The microcontroller also controls the applied field via the current through the coil integrated into the sample holder. }
\end{figure}

The original power supply inside the HV-rack used for the heating of the samples is removed and replaced by a tailor-made unit. A schematic of the total setup is shown in Fig. \ref{figure:gate-setup}.
The current pulses are produced by a commercial function generator (Agilent 33250A) and subsequently amplified by a factor of 20 up to $\pm 200$\,V. This home built amplifying stage is necessary to achieve the high current densities required for the experiment.

To drive the magnetic coil inside the sample holder, a small power supply is also integrated into the setup. Depending on the experiments, either short and strong field pulses are required to reset the magnetization and the actual experiment is done at zero field or small permanent fields are applied to set a preferred magnetization direction. The former is achieved best by an air coil without any remanent field, whereas for the latter task a coil with a mu-metal core is more suited, since it results in higher fields. The pulse generator and the coil are controlled by a microcontroller (type ATmega644 \cite{heyne}). Via a glass fiber connection all functions can be set externally using a home-made software package.  %  that can be integrated into the experiment.
By imaging before and after pulse injection, this setup is suited to study permanent changes of the magnetization induced by current injection, such as domain wall motion.

However, reversible changes of the magnetization during pulse injection cannot be studied, since the system will relax back to the stable ground state after the excitation by the current pulse. Imaging the excited state would require injection of a DC-current and simultaneous imaging. Due to the required high current densities  (about $1\cdot 10^{12}$\,A/m$^2$) and the resulting heating this is not possible. A solution is the injection of a continuous pulse train with microsecond long pulses and sufficient cooling time in between. Switching on the imaging unit only during these current pulses results in an image  of the excited state. The Joule heating is reduced by the pulse train duty cycle at the cost of an increased exposure time.

The PEEM uses a multi channel plate (MCP) to amplify the emitted photo electrons before converting them into visible light by a phosphorous screen. The MCP is typically operated at about 1500\,V. Due to the non-linear characteristic of the MCP, it is effectively switched off at voltages below 1200\,V. Thus it can be switched on and off by a gating voltage of about 300\,V on top of a constant bias voltage of 1200\,V. This fast 300\,V switch outputs 8\,ns short pulses and is controlled by a TTL-signal.
A trigger signal from the pulse generator is used by the microcontroller to generate a signal for the gating of the MCP. The timing can be set so that the imaging unit is switched on during the current pulses resulting in the excited state being imaged. Alternatively the timing can be reversed so that in between the current pulses the ground state is imaged.
Since the image is acquired over a large number of current pulses, this experiment requires reproducible states, similar to pump-probe experiments.
%\begin{figure}\center
%\includegraphics[width=\figurewidth]{figures/gatetiming.ps}  %8cm   14
%\caption[Schematic of the timing for the gating experiment]{\label{figure:GatingSchematic}Schematic overview over the timing for the gating experiment. Current pulses are injected into the sample (green pulses). For imaging of the structure during pulse injection, the imaging unit is switched on accordingly (center red line). Imaging between current injections is achieved by reversing the gating signal. (bottom red line) }
%\end{figure}

\subsection{ C. Nanosecond Pulse Injection}
\begin{figure}\center
\includegraphics[width=\figurewidth]{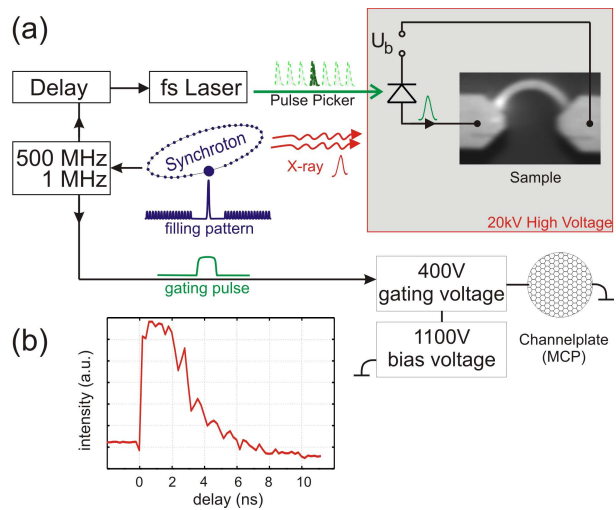}  %8cm   14
\caption[Pump-probe setup]{\label{figure:laser}(a) Setup of the laser pump-probe experiment. In the hybrid synchrotron mode, one electron bunch and the corresponding X-ray pulse is particularly intense, and the MCP is only switched on during this pulse. The laser pulses are synchronized to this pulse with a variable delay to conduct time resolved experiments. Additionally, the pulse picker can select single laser pulses. (b) Shape of the current pulse. The rise time is below 100\,ps and the pulse length is about 3\,ns}
\end{figure}

Due to the high frequency limitations of the cables connecting the PEEM with the HV-rack, nanosecond short current pulses cannot be injected into the structures
via these cables. The only solution is to directly generate such short pulses close to the sample. This is done by a fast laser pulse that triggers a photo diode mounted on the sample holder. The setup is schematically shown in Fig. \ref{figure:laser}.

The photo diode is biased by a voltage $U_b$ of about 10\,V supplied via the contacts of the sample holder (see Fig. \ref{figure:laser}(a)). The charge carriers created in the diode by the laser pulse can either be injected into a strip line below the sample to induce a magnetic field pulse or directly into the structure to study spin-torque phenomena. The laser operates at a repetition rate of 5.2\,MHz and produces pulses shorter than 50\,fs. A pulse picker can be used to select single pulses, making it possible to inject one short pulse instead of a continuous pulse train. The pulse shape is shown in Fig. \ref{figure:laser}(b). The 3\,ns short current pulse has a very fast rise time below 100 ps.

In continuous mode, the laser pulses can be synchronized to the X-ray pulses with a 500MHz signal that is supplied by the synchrotron \cite{Quitmann2008}. In normal operation, an equally intense X-ray pulse is emitted every 2\,ns, corresponding to the 500\,MHz signal. In the hybrid mode available at the SLS, a special electron bunch filling pattern is used (see Fig. \ref{figure:laser}). One out of the 480 electron bunches circulating the synchrotron is especially intense and the adjacent bunches have extremely low intensity. By gating the micro channel plate accordingly, only this
bunch will contribute to the PEEM image \cite{Quitmann2008}. Changing the delay between this X-ray beam and the laser pulse that leads to the excitation of the sample thus makes it possible to study the temporal evolution of the system after the excitation \cite{Quitmann2008, Raabe2005, Rhensius2010}.

Thus two types of experiment can be carried out: single nanosecond pulses can be injected and the resulting permanent changes of the spin structure can be imaged or time-resolved pump probe measurements can be carried out. The latter  requires a stable magnetic ground state to which the system can relax after the excitation.

\section{ III. Results}
Employing the two setups explained above allows for new experiments probing current induced magnetization manipulation. In this section one example is given for each of the setups showing the flexibility of the presented experimental scheme.

\subsection{ A. Current-Induced Vortex Core Displacement}
\begin{figure}\center
\includegraphics[width=\figurewidth]{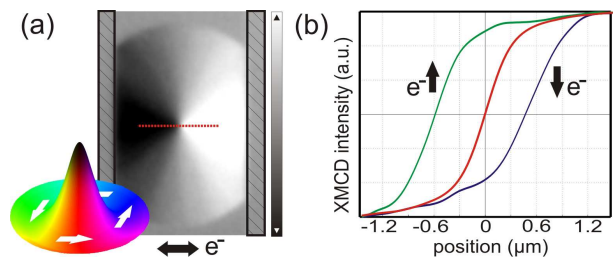}  %8cm   15
\caption[Images of vortex core displacements in permalloy disk]{\label{figure:VC_displacements} (a) X-PEEM image of the centered clockwise vortex at zero current in a 6\,$\mu$m wide and 30\,nm thick Permalloy disk. Gold contacts (shaded) to the left and to the right are used for current injection. (b) Line scans across the vortex core for the centered core (red) and for a current density of $j=+0.8 \cdot 10^{12}A/m^2$ (blue) and $j=-0.8 \cdot 10^{12}A/m^2$ (green) . The red line scan of the centered vortex was taken along red dotted line in (a).}
\end{figure}

In most disk geometries made of a soft magnetic material such as Permalloy (Ni$_{80}$Fe$_{20}$), the magnetic ground state is found to be the vortex state (see Fig. \ref{figure:VC_displacements}) . In this configuration the magnetization is curling around the center, the so called vortex core. A micromagnetic simulation of a vortex core is shown in the inset of Fig. \ref{figure:VC_displacements}(a). The color code and the arrows indicate the in-plane magnetization direction. To minimize the exchange energy, this vortex core tilts out-of-plane as visible in the simulation. This structure is quite stable and can be treated as a quasi particle \cite{Huber1982}. The study of its response to external fields and injected currents helps to gain a better understanding of the processes involved. So far the excitation of a vortex has been studied using AC magnetic fields or currents, but a study of the effect of DC currents injected into the structure was missing. It is predicted that such a study allows one to determine the size of the non-adiabaticity parameter $\beta$ \cite{Krueger2009,Moon2009}. The knowledge of this parameter is important for a better understanding of the related spin torque effect.

 This study is complicated by the fact that the required high current densities result in excessive Joule heating and therefore limit the maximum pulse length to below 1\,ms.
The setup presented above allows for injecting current pulse trains with pulse lengths of 10-100\,$\mu$s and to image the investigated structure only during these pulses. We use slow rise-times and fall-times for the pulses to prevent any dynamic gyrotropic motion and since the magnetization dynamics takes place on a nanosecond timescale, these current pulses can be considered as DC-currents.
The structure used is presented in the schematic drawing in Fig. \ref{figure:gate-setup}, where a scanning electron microscopy image of a 30 nm thick Permalloy disk with a diameter of 6 $\mu m$ is shown. The disk is contacted by Au pads for current injection at the left and right. An X-PEEM image of the magnetic initial vortex configuration with a centered vortex core is shown in Fig. \ref{figure:VC_displacements}(a). Line scans across the vortex core are shown in Fig. \ref{figure:VC_displacements}(b).
At an injected current density of about $j=\pm0.8\cdot 10^{12}$\,A/m$^2$ the original vortex (red line scan) is shifted. Reversing the current direction reverses the vortex core displacement as also expected by the theory \cite{Shibata2006}. An detailed analysis of the vortex core displacement direction will be published elsewhere~\cite{heyne2010b}.
%The required high current densities and the associated Joule heating do not allow for continuous current injection. So a continuous current pulse train is used at a low duty cycle and microsecond long current pulses. The imaging unit is only switched on during the current pulses as schematically explained in Fig. ??.
%First measurements indicate that the non-adiabaticity parameter $\beta$ is much larger than the damping constant $\alpha$ \cite{heyne2010b}, revealing that the non-adiabaticity for systems containing large magnetization gradients is on a par with the adiabatic spin-torque, which was previously believed to dominate. For theoretical models this means that spin relaxation, which predicts $\beta$ to be of the order of $\alpha$  is not sufficient but non-adiabatic transport has to be considered as a source for $\beta$ \cite{Tatara2008a}.

\subsection{ B. Current-induced Domain Wall Motion}
\begin{figure}\center
\includegraphics[width=\figurewidth]{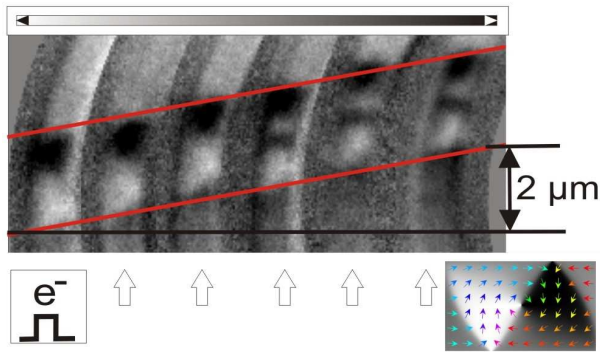}  %8cm   15
\caption[Image sequence of subsequent  current injections]{\label{figure:shortpulse}X-PEEM image series of a permalloy wire (1\,$\mu$m wide and 40\,nm thick) containing a vortex wall. The left image shows the configuration after the initialization by a short vertical field pulse. The X-PEEM images were taken with vertical contrast (see gray scale bar at the right). Current pulses were injected between adjacent images (from left to right). After the third pulse, the domain wall is transformed to a double vortex wall that continues moving.}
\end{figure}

Using the setup explained in Section IIc we have studied the current-induced domain wall motion using 3\,ns short current pulses \cite{Heyne2010}. A SEM image of the curved wire is shown in Fig. \ref{figure:laser}.
To create a domain wall prior to the current injections, an external magnetic field is applied in the vertical direction. After reducing the field to zero, a domain wall is formed at the center of the wire. An X-PEEM image of the initial configuration is presented at the left of Fig. \ref{figure:shortpulse} showing a vortex wall at the wire center. To visualize the spin structure, the inset in the lower right shows a micromagnetic simulation of a vortex wall confined in a wire. The magnetization curls around the center of the wall, the vortex core. Once the initial configuration is imaged, single current pulses are injected into the structure. The result of a series of injections is presented by the image series in Fig. \ref{figure:shortpulse}. Current pulses were injected between adjacent images (from left to right). After five injections, the domain wall is displaced by about 2\,$\mu$m. The average measured displacement per current pulse is 400\,nm. This corresponds to an average domain wall velocity of $v=130$\,m/s (3\,ns pulse length).

These high velocities are attributed to the very steep rise time of our current pulses  and the associated additional spin torque \cite{Heyne2010} and are in contrast to the low observed domain wall velocities using microsecond long current pulses \cite{Klaeui2005, Heyne2008}. In addition, domain wall transformations are observed and the resulting complicated domain walls appear to be moving with the same velocity (bottom three images in Fig. \ref{figure:shortpulse}).

\section{ IV. Conclusion}
In conclusion, we have presented a new measurement setup that allows for a study of the interaction of injected currents with the local magnetization in structured samples by using high resolution X-PEEM imaging.
Stable magnetic configurations such as magnetic vortices in disks can be excited by a current and reversibly change their configuration. For instance an injected current displaces the vortex core, but after the current is switched off, the core relaxes back to its initial position. Due to the required high current densities and the excessive Joule heating, the imaging of this excited state  which has note been possible previously. By using only microsecond long current pulses and gating the imaging unit accordingly, we succeeded in imaging the reversible current-induced vortex core displacement. This yields important information about the relative strengths of the spin torque terms.

The in-situ injection of single nanosecond short current pulses is made possible by the use of a fast photo diode in combination with a fs-laser and a specially designed sample holder. The current pulses are created close to the sample and have a steep rise time of about 100\,ps. This setup has been used successfully to study the current-induced domain wall motion in permalloy wires, which highlights the importance of the pulse shape for the resulting domain wall velocities.
The setup has reached a stage of maturity where it is user friendly, reliable while retaining its flexibility to study various aspects of the interaction of spin-polarized currents and the magnetization in photoemission  electron microscopes. Future developments could include a lock-in amplifier to measure the resistance with high precision in-situ and replacing the contacts with high frequency cables so that an even more flexible setup results, which would open up new opportunities and eventually become a user facility at synchrotron facilities.\\
\section{Acknowledgments}
Support by F. Kronast is acknowledged for the design of the sample holder and J. Raabe for the design of the MCP gating. We thank A. Locatelli from the ELETTRA synchrotron and S. Dhesi from the DIAMOND synchrotron for fruitful discussions. The authors further acknowledge financial support by the German Science Foundation (SFB 767, KL1811), the European Research Council (Grant No. ERC-2007-Stg 208162), the European Community under the 7th Framework Program  (Grant No. NMP3-SL-2008-214469) and the German National Academic Foundation (LH). Part of this work was carried out at the Swiss Light Source and at BESSY.

\end{document}